**Comment on "On the Yield of Singlet Excitons in Organic Light-Emitting Devices: a Double Modulation Photoluminescence-Detected Magnetic Resonance Study"**

C. G. Yang[1], E. Eherenfreund[2] and Z. V. Vardeny[1,2]
[1]Physics Department, University of Utah, Salt Lake City, Utah 84112, USA
[2]Physics Department and Solid State Institute, Technion, Haifa 32000, Israel

In a recent Letter [1] Lee *et al.* studied the spin ½ photoluminescence-detected magnetic resonance (PLDMR) of a soluble derivative of poly-phenylene vinylene [MEH-PPV] using a double modulation (DM) scheme, where both the laser excitation intensity and the microwave power were modulated at $\omega_{ex}$ and $\omega_\mu$ ($\omega_\mu \ll \omega_{ex}$), respectively. The obtained results were used to attempt differentiating between the quenching and the spin dependent recombination (SDR) models for PLDMR [2]. Also, an alternative model to the SDR was introduced [1] in which triplet excitons interact with polarons, (T-P model) leading to a reduction of both species. It was concluded [1] from the $\omega_\mu$-dependence of PLDMR that polaron recombination time is ~30 μs. This conclusion is unjustified, however since the authors overlooked (already in Eq. (1)) the role of polaron lifetime distribution, and spin subsystem dynamics on the PLDMR frequency response. In addition, the T-P model [1] is not viable in π-conjugated polymers in general, and in MEH-PPV in particular.

In textbook introduction to PLDMR [3] the spin-lattice relaxation and microwave (μw) power are shown to play an important role in determining the PLDMR dynamics and magnitude, especially for SDR [4]. Since SDR is inherently dispersive with a very broad distribution of recombination times τ [3], then the spin-lattice relaxation time, $T_{SL}$ separates pairs with $\tau \ll T_{SL}$ that contribute strongly to PLDMR [5] from those pairs with $\tau \gg T_{SL}$ that contribute very weakly to PLDMR [3-5]. Thus the $\omega_\mu$-dependence PLDMR is sensitive to the *spin dynamics* as well as recombination kinetics of the spin carrying excitations.

We have compared the $\omega_{ex}$-dependence of polarons in MEH-PPV at 10 K using the photoinduced absorption (PA) technique at 0.4 eV with the spin ½ PLDMR $\omega_\mu$-dependence at different μw powers (Fig. 1). These modulation frequency dependencies are not similar to each other for both in-phase and quadrature components, thus revoking the claim in [1] that the spin ½ PLDMR $\omega_\mu$-dependence "yields the lifetimes of the spin-carrying species responsible for the resonance". In particular, it is clear from Fig. 1 that the recombination mechanism of the polaron PA, unlike the PLDMR, is both *slower* and *dispersive* due to a very broad distribution of lifetimes. In addition, the PLDMR response depends on the μw power, whereas the PA response does not. We thus conclude that the apparent time-constant in the PLDMR $\omega_\mu$-dependence *is not* the spin-carrying recombination lifetime as claimed in [1]; but instead is determined by $T_{SL}$, μw power, and τ distribution. Consequently, the analysis of the flat DM $\omega_{ex}$-dependence in [1] that is based on Eq. (1) is questionable.

Second, the T-P model advanced in [1] cannot explain the existence of spin ½ PLDMR when the triplet exciton density is negligibly small. This occurs in MEH-PPV at elevated temperatures (~230K) where the triplet PA is negligible whereas the spin ½ PLDMR dynamics does not change much; as well as in many π-conjugated polymers where the triplets are unstable. Two examples of such polymers include: (i) $C_{60}$-doped MEH-PPV, in which the triplet excitons dissociate at the $C_{60}$ sites [6]; yet we measured $\delta L/L \sim 4 \times 10^{-4}$ at $g \cong 2$. (ii) The degenerate ground state polymer poly(di-phenyl-acetylene), in which the triplet excitons are unstable against dissociation into solitons pairs [7]; yet we measured $\delta L/L \sim 2 \times 10^{-4}$. The underlying mechanism of the spin ½ PLDMR in these examples *cannot be associated with triplet excitons* and thus the SDR model, which does not explicitly involve triplets [2-5] can uniquely explain the results. Consequently, if the SDR model is viable when the triplet density is small, then this mechanism cannot be turned off in other situations and polymers.

[1] M. –K. Lee *et al.,* Phys. Rev. Lett. **94,** 137403 (2005).
[2] M. Wohlgenannt *et al.,* Nature **409,** 494 (2001).

[3] B. C. Cavenett, Advances in Physics **30**, 475 (1981).
[4] E. Lifshitz *et al.,* Annu. Rev. Phys. Chem, **55**, 509 (2004).
[5] R. A. Street, D. K. Biegelsen, and J. Zesch, Phys. Rev. B **25**, 4334 (1982).
[6] X. Wei, *et al.*, Phys. Rev. B **53,** 2187 (1996).
[7] I. I. Gontia, *et al.*, Phys. Rev. Lett. **82**, 4058 (1999).

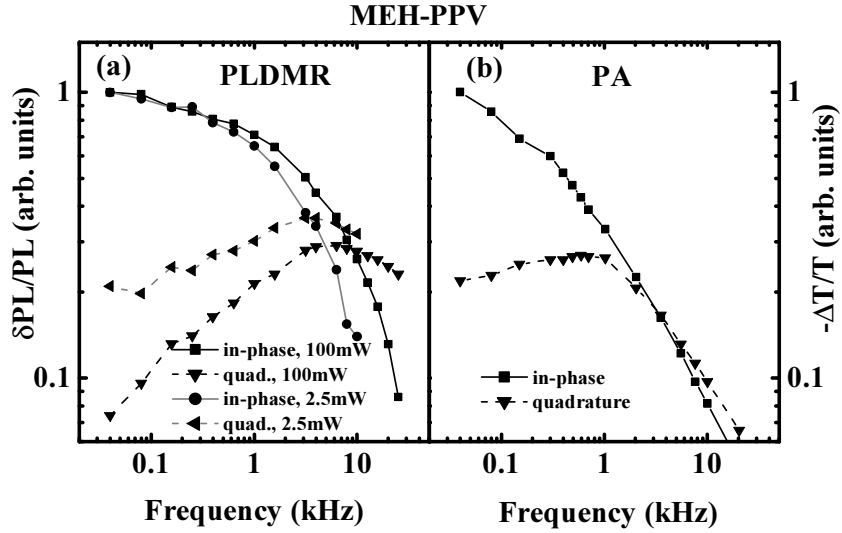

FIG. 1: (a) The μw modulation frequency dependence of spin ½ PLDMR in MEH-PPV film measured at 10 K at two μw powers. (b) The laser excitation frequency dependence of the polaron PA at 0.4 eV. The PA response is dispersive and does not depend on the μw power.